\documentclass[12pt,preprint]{emulateapj}
    \usepackage{apjfonts,psfig}
    \slugcomment{ApJ, accepted for publication}
    \received{}
    \revised{}
    \accepted{}
    
\shortauthors{Remco van den Bosch et al.}
\shorttitle{Dynamics of M15}

\newcommand{\pc}{\>{\rm pc}}
\newcommand{\Msun}{\>{\rm M_{\odot}}}
\def\kms{km\,s$^{-1}$}

\def\masyr{mas\,yr$^{-1}$}

\def\farcm{\hbox{$.\mkern-4mu^\prime$}}
\def\farcs{\hbox{$.\!\!^{\prime\prime}$}}

\newcommand{\figdisp}
{Smoothed radial dispersion profiles of the line-of-sight velocities
and proper motions for various cuts in the velocity error
distribution.  a) Line-of-sight velocities. Red is an uncertainty cut
of 5 \kms\ or less, black is for 7 \kms, and blue is for 14 \kms.
Uncertainty cuts below 5 \kms\ show no difference from the red
lines. Every error bar represents a radial ring containing 60
stars. b) major (solid lines) and minor (dashed lines) axis proper 
motions for various cuts in the
velocity error distribution. Red is an uncertainty cut of 8 \kms\ or
less, black is for 14 \kms, and blue is for 21 \kms (for an assumed
distance of 10 kpc).  Uncertainty cuts below 8 \kms\ show no
difference from the red curves. Every error bar represents a radial
ring containing approximately 60 stars. \label{fig:disp}}

\newcommand{\figmlosv}
{Smoothed line-of-sight mean velocity (top) and velocity dispersion
field (bottom) for M15 in \kms. \label{fig:mlosv}}

\newcommand{\figmpmvel}
{Smoothed kinematic velocity fields of the proper motions in \kms\
assuming a distance of 10 kpc. {\em Left:} kinematics in the
$x'$-direction (parallel the major axis). {\em Right:} kinematics in
the $y'$-direction (parallel the minor axis). {\em Top:} Velocity
field. {\em Bottom:} Corresponding dispersion
fields. \label{fig:mpmvel}}

\newcommand{\figmge}{{\em Top:} Radial surface-brightness profile of M15
from Noyola \& Gebhardt (2005) and the corresponding MGE best-fit
model consisting of 14 Gaussians as a function of radius. Each
individual Gaussian is shown as well. Left axis shows L$_{\odot}$ per
pc$^{2}$. {\em Bottom:} The residual between the surface brighness and
the MGE best-fit model. The profile is reproduced to within 3\% over
the entire radial range. \label{fig:mge}}

\newcommand{\figkinbestfit}
{Observed mean velocity and velocity dispersion (first and third
column) and the corresponding kinematics from the best-fit dynamical
model with $D=10.3$ kpc (second and fourth column). Top and middle row
show the average kinematics of the proper motions in the $x'$ and
$y'$-direction and the bottom row shows the averaged kinematics of the
line-of-sight velocities. \label{fig:kinbestfit}}

\newcommand{\figchiincdist}
{Marginalized $\chi^2$ contour map of the models (crosses) varying
inclination (horizontal axis) and distance (vertical). The three inner
contours are drawn at formal 68.3\%, 95.4\% and 99.7\% confidence
levels for one degrees or freedom. The subsequent contour corresponds to
a factor two increase in $\Delta\chi^2$. The contours outside the models
are extrapolated. The best-fit model at inclination 60$^\circ$ and
10.3 kpc is denoted with a black star. The inclination is not well
constrained. \label{fig:chiincdist}}

\newcommand{\figmlprof}
{Radial $M/L$ Profiles. {\em Black line:} Our best-fit V-band MGE
$M/L$ values and error bars signifying 68\% confidence for five degrees
of freedom. {\em Red Line} Our best-fit deprojected $M/L$
profile. {\em Blue solid line:} The deprojected $M/L$ profile from
Baumgardt (priv. comm.) {\em Blue dashed line:} Profile from Pasquali
et al. See Section \ref{sec:mlprofile} and \ref{sec:mldcm} for a
discussion.
\label{figmlprof}}

\newcommand{\figchibhcnt}
{Marginalized $\chi^2$ contour map of the models (crosses) varying the
central M$_{\odot}$/L$_{\odot}$ value (horizontal) and the dark
central mass (M$_\odot$, vertical axis). The three inner contours are
drawn at formal 68.3\%, 95.4\% and 99.7\% confidence levels for three
degrees of freedom. Subsequent contours correspond to a factor 2
increase in $\Delta\chi^2$. The best-fit model with a dark central
mass of 500 M$_{\odot}$ and a central $M/L$ of 5
M$_{\odot}$/L$_{\odot}$ is denoted with a black star. The correlation
between the two parameters can clearly be seen, showing that the data
primarily constrains the total central mass enclosed.
\label{fig:chibhcnt}}

\newcommand{\figimage}{Images of the central region of M15, $9''$ on
a side. The left image is a WFPC2/F336W exposure. The right image is
heavily massaged; we have convolved the original image with a boxcar
of 20 pixels in size. The blue curves are smoothed isophotes at
$1\farcs5$, $2\farcs2$ and $2\farcs8$ major axis radius. The straight
line is the best-fit position angle for the isophotal major
axis. \label{fig:image}}

\newcommand{\figrotpmrv}{Velocity of stars versus position angle on the
sky for each velocity component. The top is for the motion parallel to
the major axis, the middle is along the minor axis, and the bottom is
the line-of-sight motion. The line is the best fit profile with a
three-parameter fit. We fit the position angle of the major axis,
amplitude of the rotation along minor and line-of-sight axes, and
amplitude along the major axis. The position angle of the maximum in
the line-of-sight velocity corresponds to the isophotal major axis of
the central structure in M15.
\label{fig:rotpmrv}}

\begin{document}

\title{The dynamical $M/L$-profile and distance of the globular 
       cluster M15.\altaffilmark{1}}

\author{%
Remco\ van den Bosch\altaffilmark{2}, Tim de
Zeeuw\altaffilmark{2}, Karl Gebhardt\altaffilmark{3}, \\ Eva
Noyola\altaffilmark{3}, Glenn van de Ven\altaffilmark{2} }

\email{bosch@strw.leidenuniv.nl}

\altaffiltext{1}{Based on observations with the NASA/ESA Hubble Space
                 Telescope obtained at the Space Telescope Science
                 Institute, which is operated by the Association of
                 Universities for Research in Astronomy, Incorporated,
                 under NASA contract NAS5--26555.}

\altaffiltext{2}{Sterrewacht Leiden, Postbus 9513, 2300 RA Leiden,
                 The Netherlands.}

\altaffiltext{3}{Astronomy Department, University of Texas, Austin.}

\begin{abstract}
We construct orbit-based axisymmetric dynamical models for the
globular cluster M15 which fit groundbased line-of-sight velocities
and Hubble Space Telescope line-of-sight velocities and proper
motions. This allows us to constrain the variation of the
mass-to-light ratio $M/L$ as a function of radius in the cluster, and
to measure the distance and inclination of the cluster. We obtain a
best-fitting inclination of $60^\circ\pm15^\circ$, a dynamical
distance of $10.3\pm0.4$ kpc and an $M/L$ profile with a central
peak. The inferred mass in the central 0.05 parsec is 3400 $\Msun$,
implying a central density of at least $7.4\times10^6\Msun$
pc$^{-3}$. We cannot distinguish the nature of the central mass
  concentration. It could be an IMBH or it could be large number of
  compact objects, or it could be a combination. The central 4 arcsec of M15 appears to contain a rapidly
spinning core, and we speculate on its origin.
\end{abstract}


\section{Introduction}
\label{sec:intro}

M15 is a well-studied globular cluster. It has a very steep central
luminosity profile, and may be in the post-core-collapse stage (e.g.,
Phinney 1993; Trager, King \& Djorgovski 1995).  Measurements of
nearly two thousand line-of-sight velocities (from the ground and with
HST) and proper motions (with HST) have recently become available
(Gebhardt et al.\ 2000, hereafter G00; McNamara, Harrison \& Anderson
2003, hereafter M03).

McNamara, Harrison \& Baumgardt (2004, hereafter M04) restricted
themselves to the subset of 237 stars inside $0\farcm3$ of the center
of M15 for which both Fabry--Perot radial velocities and HST proper
motions were measured, and computed the mean dispersions in these
measurements. Assuming the cluster is an isotropic sphere, and the
observed stars are representative, the ratio of these dispersions (one
in \kms, the other in \masyr) provides the distance (Cudworth 1979,
Binney \& Tremaine 1987). M04 find a distance of $9.98\pm0.47$ kpc,
which is consistent with the canonical value of 10.4 kpc (Durrell \&
Harris 1993), but is smaller than, e.g., the recent determination of
11.2 kpc by Kraft \& Ivans (2003) who used a globular cluster
metallicity scale, based upon Fe~II lines.

Here we extend the M04 study by using a larger fraction of the
line-of-sight velocity and proper motion samples, and comparing these
with more general dynamical models to study the internal structure of
the cluster as a function of radius. We follow the approach taken by
van de Ven et al.\ (2005, herafter V05), who constructed axisymmetric
dynamical models for the globular cluster $\omega$ Centauri and fitted
these to groundbased proper motions and line-of-sight velocities. This
technique provides the internal dynamical structure as well as the
inclination of the cluster, an unbiased and accurate dynamical
distance, and the $M/L$-profile. Our aim is to derive similar
information for M15. We are particularly interested in the $M/L$
profile, as significant mass segregation is believed to have occurred
in the cluster (Dull et al.\ 1997). The HST proper motions have
sufficient spatial resolution to study the dynamical structure and
mass concentration in the center.

In Section~\ref{sec:data}, we summarize the observational data. In
Section~\ref{sec:second-moments}, we consider the influence of
measurement errors on the data, select the stars to be used for the
dynamical modeling and also study the possible residual systematic
effects in the observed mean motions. We construct dynamical models in
Section~\ref{sec:models}, derive a distance, and investigate the
effect of the unknown inclination and of radial $M/L$ variations in
the cluster. We discuss the dynamics of the central 0.2 parsec of M15
in Section~\ref{sec:spinning-core}, and summarize our conclusions in
Section~\ref{sec:discussion}.

\begin{figure*}
\epsscale{1.0}
\plotone{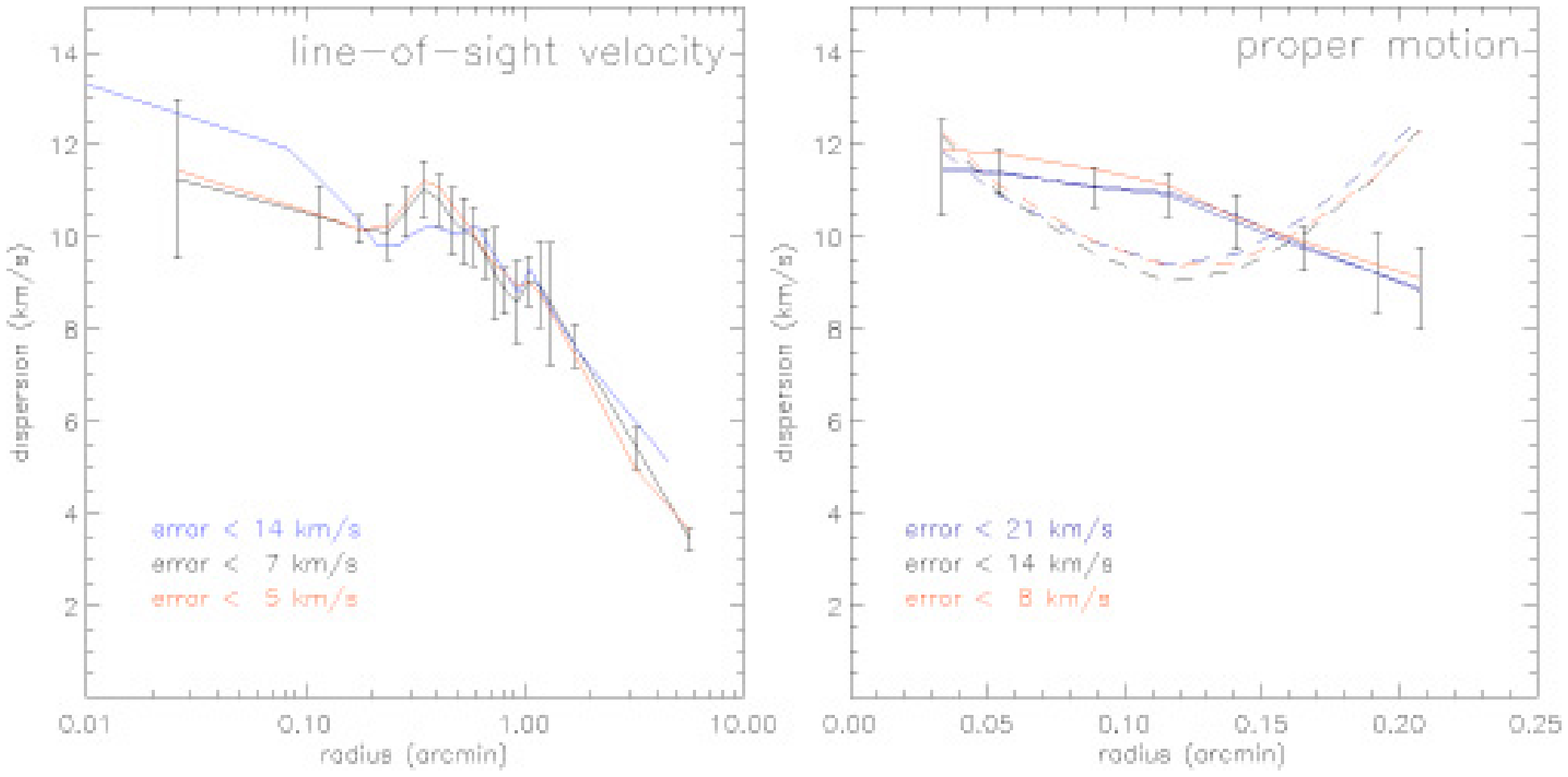}
\figcaption{\figdisp}
\end{figure*}

\section{Observational Data}
\label{sec:data}

We discuss, in turn, the surface brightness distribution, the
line-of-sight velocities, and the proper motion data for M15.

\subsection{Surface brightness distribution}
\label{sec:surfbrightdist}

The surface brightness distribution of M15 has been studied in detail
by a number of authors (Lauer et al.\ 1991; Trager et al.\ 1995;
Guhathakurta et al.\ 1996; Sosin \& King 1997). Noyola \& Gebhardt
(2005) reanalysed archival WFPC2 images with a new technique which
measures the integrated light to determine the surface-brightness
distribution in the dense central regions of the cluster. They
combined this with the groundbased profile from Trager et al.\
(1995). The surface-brightness profile is a power-law in the inner
arcsecond, with slope $-0.62\pm0.06$, which agrees with the
determinations by Guhathakurta et al.\ (1996) and Sosin
\& King (1997).\looseness=-2

Because the surface-brightness distribution of M15 is only slightly
flattened and the individual stars in the cluster are resolved, it is
difficult to estimate the ellipticity and the position angle (PA) of
the major axis. Guhathakurta et al.\ (1996) measured an average
ellipticity of $\epsilon = 0.05\pm0.04$.  Determinations of the PA of
the photometric major axis as measured from North through East vary
from $125^\circ$ (White \& Shawl 1987) to $45^\circ$ (from a DSS
image) at large radii, and is given as $60^\circ\pm20^\circ$ by
Guhathakurta et al.\ (1996). Gebhardt et al.\ (1997) reported the
kinematic PA of maximum rotation at $198^\circ$ North through East.
Here we adopt this value for the PA of the major axis (see
Section~\ref{sec:lummodel}).

\subsection{Line-of-sight velocity sample}

The bulk of the line-of-sight velocities come from the compilation by
G00, who reported measurements for 1773 stars brighter than B
magnitude 16.5 out to a radius of $17'$. Their data set includes
earlier measurements by Peterson, Seitzer \& Cudworth (1989), Dubath
\& Meylan (1994), Gebhardt et al.\ (1994, 1995, 1997) and Drukier et
al.\ (1998), as well as 82 stars in the inner region measured using
adaptive-optics-assisted spectroscopy (G00). In addition,
line-of-sight velocities for 64 stars in the inner 4$''$ were obtained
with STIS onboard HST (van der Marel et al.\ 2002; Gerssen et al.\
2002, 2003).

The expected number of non-members in this data set is very small. The
cluster is very dense, so few interlopers are expected from chance
superposition of field stars. The systemic line-of-sight velocity of
M15 is $-107.5\pm0.2$ \kms\ (G00). With an internal velocity
dispersion of about 12 \kms\ in the center, the cluster stars are
well-separated in velocity from the foreground galactic disk.  The
bulk of the data is inside $4'$ and the median error of the
line-of-sight velocities is 3.5 \kms. This is a significant fraction
of the line-of-sight velocity dispersion, especially at larger radii.

\subsection{Proper motions}

M03 published proper motions for 1764 stars within $0\farcm3$ of the
center of M15, derived from multi-epoch HST/WFPC2 imaging. The stars
range in brightness between 14.0 and 18.3 mag. Of the 1764 stars only
703 stars brighter than $B$-magnitude 16.5 where kept and used in
their analysis. We restrict ourselves to this sample of 703 stars.

M03 derived the proper motions in the classical way, by using a
reference frame consisting of cluster stars, and then modeling the
difference between the positions on the first and second epoch images
in terms of a zero point difference, a scale change, a rotation, a
tilt, and second-order distortion corrections (e.g., Vasilevskis et
al.\ 1979).  The corresponding transformation equations were then
solved using a least squares routine. As a result, the derived proper
motions are not absolute, but may contain a residual global rotation.
They also may be influenced by perspective rotation caused by the
space motion of the cluster. We return to this in
Section~\ref{sec:mean-motions}.

Figure 1 of M04 shows the histogram of all proper motions. The median
error of these measurements is 0.12 mas yr$^{-1}$, corresponding to
about 6 \kms\ at a distance of 10 kpc, with some errors as large as
0.50 mas yr$^{-1}$.

\section{Kinematics}
\label{sec:second-moments}
\label{sec:mean-motions}

We use a Cartesian coordinate system $(x', y', z')$ with $z'$ along
the line of sight and $x'$ and $y'$ in the plane of the sky aligned
with the cluster such that the $y'$-axis is the photometric minor and
rotation axis of the cluster. The kinematic measurements then give
$v_{z'}$ for the line-of-sight velocities in \kms, and $\mu_{x'}$ and
$\mu_{y'}$ for the proper motions in \masyr. To convert $\mu_{x'}$ and
$\mu_{y'}$ into $v_{x'}$ and $v_{y'}$ in \kms, we used
$v_{x'}=4.74D\mu_{x'}$ and $v_{y'}=4.74D\mu_{y'}$, with $D$ the
distance in kpc.

The measurement errors in the line-of-sight velocities and in the
proper motions need to be taken into account when
analyzing the kinematics of M15. We do this by means of a maximum
likelihood method, which corrects for each individual velocity error,
and provides robust estimates of the mean velocities and velocity
dispersions in spatial bins on the sky. The method is described in
detail in Appendix A of V05.

\subsection{Selection}

Figure~\ref{fig:disp}a shows the velocity dispersion along the line
of sight, as a function of radius, for three different selections of
stars. The blue curve is for all stars with velocity error smaller
than 14 \kms, the black curve for those with errors smaller than 7
\kms, and the red curve for the stars with errors smaller than 5
\kms. Although we corrected the dispersion for the individual
measurement errors, the curves do not all overlap, suggesting that the
(larger) errors are not estimated very accurately. The curves converge
once we exclude the stars with errors larger than 7 \kms, so we
restrict ourselves to this sample of 1546 stars. The corresponding
profile varies between approximately 12 \kms\ in the center to 3
\kms\ at about 10$'$ (cf.\ Figure 12 in G00).

Figure~\ref{fig:disp}b shows the velocity dispersions of the proper
motions in the $x'$ and $y'$ direction, respectively, as a function of
radius, for three different selections of stars based on the
measurement errors, corresponding to 21, 14 and 8 \kms\ for an assumed
distance of 10 kpc. The radial range covered is only $0\farcm23$, and
hence corresponds to the inner data points of Figure~\ref{fig:disp}a.
The profiles for the different error selections are consistent to
within the errors of the dispersions. The dispersion
profiles in the two orthogonal directions appear to differ
and they are, in fact, inconsistent at the formal one sigma
level. The overall proper motion dispersion is independent of the
error selection. It is therefore not evident that a selection
based on measurement error is justified, and we use all 703 proper
motions.

\begin{figure}
\epsscale{1.0}
\plotone{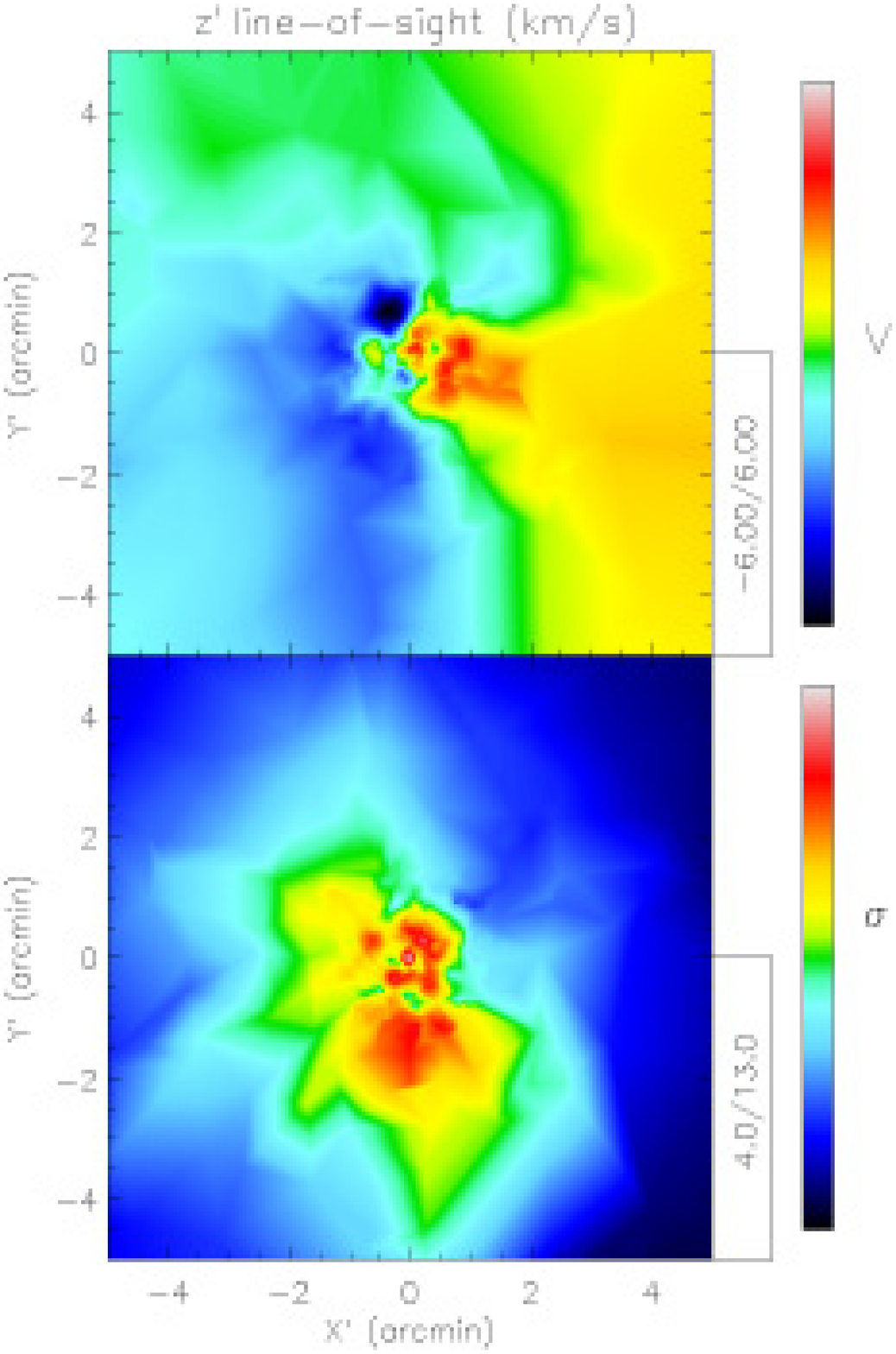}
\figcaption{\figmlosv}
\end{figure}

\begin{figure}
\epsscale{1.0}
\plotone{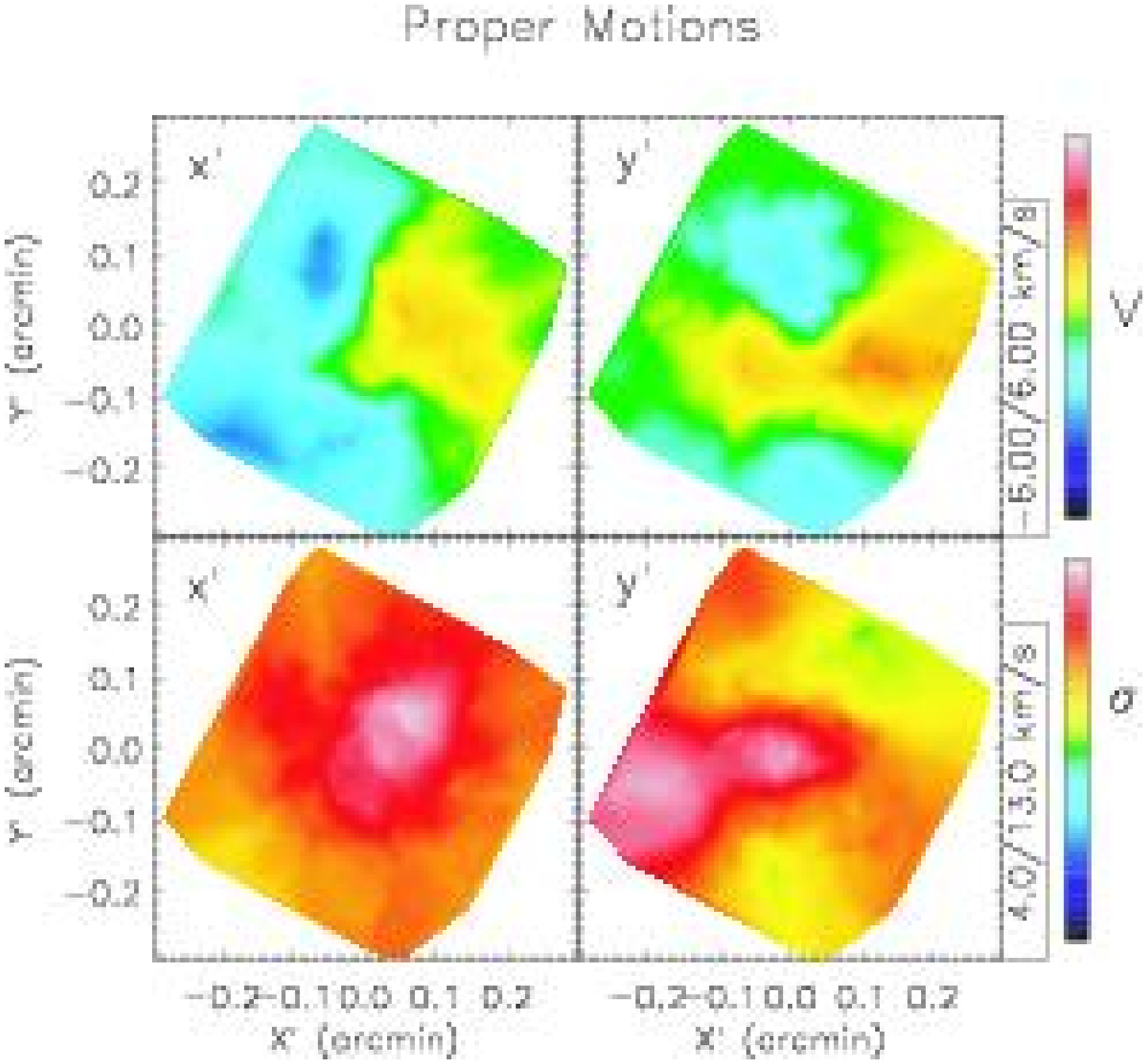}
\figcaption{\figmpmvel}
\end{figure}

Figures~\ref{fig:mlosv} and \ref{fig:mpmvel} show a smooth
representation of the mean velocity and dispersion fields of the
line-of-sight velocities and proper motions, respectively. The fields
where adaptively smoothed by computing at each stellar position the
kinematic moment for the nearest 100 neighbors using Gaussian
weighting with distance from the central star. The Gaussian used for
the weighting has the mean distance of the 100 stars as its
dispersion. The resulting smooth kinematic maps correlate the
different values at different points, but bring out the main features
of the observed kinematics of M15.

The line-of-sight velocity maps show significant structure, including
overall rotation (G00; and Section~\ref{sec:spinning-core} below).
The line-of-sight velocity dispersion shows the radial fall-off
illustrated in Figure~\ref{fig:disp}a. The contours of constant
dispersion are slightly elongated.  The mean proper motion maps are
difficult to interpret, because of the relatively large measurement
errors. The dispersions in the proper motions are nearly constant over
the small extent of the field (cf Figure~\ref{fig:disp}b).

\subsection{Perspective rotation}

\begin{deluxetable}{rrl}
\tablewidth{8truecm}
\tablecaption{Systemic proper motion of M15\label{tab:space_motion}}
\tablehead{
\colhead{$\mu_\alpha \cos\delta$} & \colhead{$\mu_\delta$}
                                   & \colhead{Reference}\\
\colhead{(1)} & \colhead{(2)} & \colhead{(3)}  \\[-10pt]
}
\startdata
  $-0.3\pm1.0$   & $-4.2\pm1.0$    &Cudworth \& Hanson (1993) \\
  $-1.0\pm1.4$   & $-10.2\pm1.4$   &Geffert et al.\ (1993) \\
  $-0.1\pm0.4$   & $0.2\pm0.3$     &Scholz et al.\ (1996) \\
  $-2.4\pm1.0$   & $-8.3\pm1.0$    &Odenkirchen et al.\ (1997) \\
  $-0.95\pm0.51$ & $-5.63\pm0.50$  &Dinescu et al.\ (1999) \\
\enddata
\tablecomments{Five determinations of the systemic motion
of M15 on the plane of the sky.  Cols (1) \& (2): Systemic proper
motion in $\alpha, \delta$ in units of mas~yr$^{-1}$. Col.\ (3):
Reference.}
\end{deluxetable}

The observed motions contain a contribution from the perspective
rotation caused by the space motion of M15.  The systemic
line-of-sight velocity is $-107.5\pm0.2$ \kms\ (Gebhardt et al.\
1997), but the component of the space motion in the plane of the sky
is not well-determined. Table~\ref{tab:space_motion} presents a
summary of the reported space motions for M15.

Use of eq.\ (6) of V05 shows that any of the values in
Table~\ref{tab:space_motion} result in contributions to the observed
proper motions of at most 0.0025 mas yr$^{-1}$ at the edge of the
small area where we have measurements.  This is well below the
measurement errors, and we therefore ignore it. The contribution to
the observed line-of-sight velocities is $\pm$0.1 \kms\ at $5'$ from
the center if we use the Cudworth \& Hanson (1993) value for the space
motion, which is small enough that it can be ignored. If the more
recent determinations by Odenkirchen et al.\ (1997) and Dinescu et
al.\ (1999) are correct, then there would be a contribution of
$\pm0.8$ \kms. This contribution is still not significant relative to
the measurement errors and therefore we do not apply any correction
for the perspective rotation. However, in this case perspective
rotation would become important for studies of the kinematics near the
tidal radius (21\farcm5, Trager et al.\ 1995). If not corrected for,
it would result in an apparent leveling-off of the line-of-sight
velocity dispersion, as reported by Drukier et al.\ (1998).

\subsection{Residual global rotation}
\label{sec:glob-rot}

It is possible to correct for the possible presence of residual global
rotation in proper motion measurements of a globular cluster by using
the line-of-sight velocities and the assumption of axisymmetry. V05
applied this method with success to $\omega$ Centauri. We follow the
same approach for M15.

In an axisymmetric cluster, the following relation is valid between
the mean motion $\langle \mu_{y'} \rangle$ and the mean line-of-sight
velocity $\langle v_{z'} \rangle$ at any point $(x', y')$:
\begin{equation}
\langle v_{z'}\rangle = 4.74 D \tan i \langle \mu_{y'}\rangle ,
\label{e:dtani}
\end{equation}
where $\langle v_{z'}\rangle$ is in units of \kms, $D$ is the distance
in kpc, $i$ is the inclination of the cluster, and
$\langle\mu_{y'}\rangle$ is in mas\,yr$^{-1}$ (cf.\ Evans \& de Zeeuw
1994). We computed the mean value  $\langle\mu_{y'}\rangle$ in
spatial bins on the plane of the sky, and similarly for $\langle
v_{z'}\rangle$. We find a formal best-fit value which corresponds to
an inclination of 59$^\circ\pm$12$^\circ$ at a distance of
10$\pm$0.5 kpc.

If the proper motion measurements still contain an unknown amount of
global rotation $\Omega$ (constant with radius), then the observed
$\mu_{x'}$ and $\mu_{y'}$ should be replaced by
\begin{equation}
\mu_{x'} =\mu_{x'} + y' \Omega, \qquad \mu_{y'} = \mu_{y'} - x'   
\Omega.
\end{equation}
We stepped through a range of values of $\Omega$ and determined the
value for which the scatter around the relation (\ref{e:dtani}) was
minimized. We find a weak minimum for $\Omega$ equal to $-0.21\pm0.22$
mas (yr arcmin)$^{-1}$. The difference between the two cases is modest
and changes the formal best-fit value of the inclination slightly to
54$^\circ\pm$10$^\circ$. We conclude that any effect of residual
global rotation is below the measurement errors, and therefore we do
not correct for it.

\section{Dynamical models}
\label{sec:models}

We construct axisymmetric dynamical models of M15 by means of
Schwarzschild's (1979) orbit superposition method, as implemented by
Verolme et al.\ (2002). The inclusion of proper motion data is
described in detail by V05, together with extensive tests designed to
establish the accuracy with which the distance and internal structure
can be recovered. These dynamical models are collisionless, which is
not necessarily a valid assumption for a dense globular cluster such
as M15. \looseness=-2

We start by constructing a luminosity model
(Section~\ref{sec:lummodel}), and compute constraints from the
observed kinematics binned into polar apertures
(Section~\ref{sec:apbinning}). In each aperture, we compare the mean
velocities and velocity dispersions to the predictions of the
dynamical model while varying the parameters to find a best-fit model.
The model parameters are the inclination $i$, the distance $D$, the
mass-to-light ratio ($M/L$) values in different radial bins and a
central dark mass $M_{\rm dark}$ (Section~\ref{sec:parameters}).  In
Section~\ref{sec:bestfit}, we obtain the best-fit models and we
discuss the results in Sections~\ref{sec:incdist},
\ref{sec:mlprofile} and \ref{sec:mldcm}.

\subsection{Luminosity model}
\label{sec:lummodel}

\begin{figure}
\epsscale{1.0}
\plotone{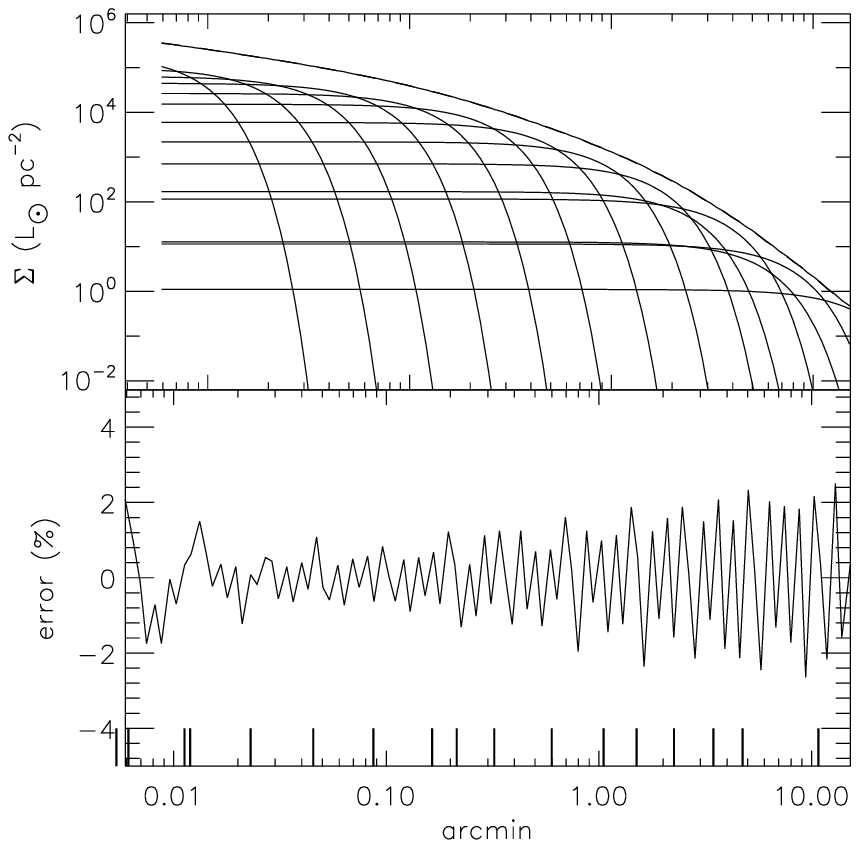}
\figcaption{\figmge}
\end{figure}

\begin{deluxetable}{rlll}
\tablewidth{5.0truecm}
\tablecaption{MGE parameters for M15\label{tab:mge}}
\tablehead{
\colhead{\#} & \colhead{log$I_{0,V}(L_\odot {\rm pc}^{-2})$} &
\colhead{log $\sigma '$(arcmin)} & \colhead{$M/L$}  \\
\colhead{(1)} & \colhead{(2)} & \colhead{(3)} & \colhead{(4)} \\[-10pt]
}
\startdata
1  &     5.306  &  -2.270  &       5.0$^{+7.0}_{-4.0}$  \\
2  &     5.009  &  -1.923  &       1.9  \\
3  &     4.822  &  -1.639  &       1.0$^{+0.9}_{-0.5}$ \\
4  &     4.670  &  -1.343  &       1.1  \\
5  &     4.442  &  -1.061  &       1.3  \\
6  &     4.202  &  -0.784  &       1.4$^{+0.4}_{-0.4}$ \\
7  &     3.795  &  -0.492  &       1.7  \\
8  &     3.361  &  -0.223  &       2.0  \\
9  &     2.871  &   0.022  &       2.3  \\
10 &     2.250  &   0.177  &       2.5$^{+0.5}_{-0.5}$ \\
11 &     2.084  &   0.353  &       2.5  \\
12 &     1.127  &   0.538  &       2.5  \\
13 &     1.080  &   0.675  &       2.5  \\
14 &     0.066  &   1.031  &       2.5  \\
\enddata
\tablecomments{
The parameters of the 15 Gaussians from the MGE-fit to the $V$-band
surface brightness profile of Noyola \& Gebhardt (2005).  Col.\ (1):
number of Gaussian component. Col.\ (2): Central surface brightness of
each Gaussian adjusted for the assumed ellipticity of
$\epsilon=0.05$. Col.\ (3) Dispersion along the major axis. Col.\ (4)
Best-fit $M_\odot/L_{\odot,V}$ value for each Gaussian and associated
error (See Section~\ref{sec:mlprofile}). }
\end{deluxetable}

\begin{figure*}
\epsscale{1.0}
\plotone{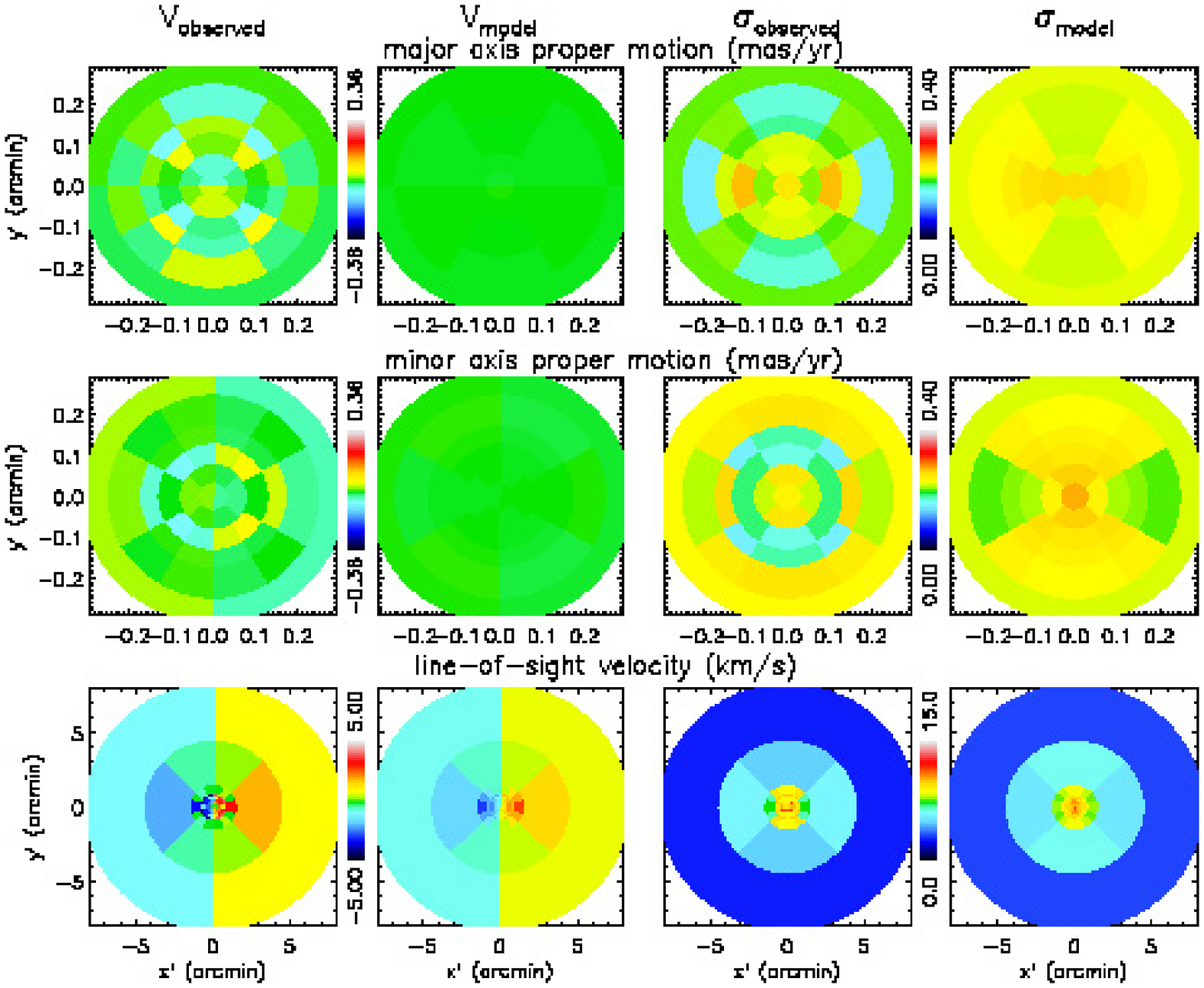}
\figcaption{\figkinbestfit}
\end{figure*}

We use the Noyola \& Gebhardt (2005) $V$-band one-dimensional
surface-brightness profile discussed in
Section~\ref{sec:surfbrightdist} as a basis for our mass model of
M15. The profile extends to 15$'$. We parameterized the profile with a
multi-Gaussian expansion method (Monnet, Bacon \& Emsellem 1992;
Emsellem, Monnet \& Bacon 1994) by means of the MGE fitting software
developed by Cappellari (2002). Figure~\ref{fig:mge} shows the
comparison between the observed profile and the MGE fit.
Table~\ref{tab:mge} gives the numerical values of the Gaussians that
comprise the MGE model. The fit is accurate to better than 3\% at all
radii. The smallest Gaussian in the MGE model has a sigma of 0\farcs23
and the largest Gaussian has a sigma of 10\farcm7. The mass outside the tidal radius is negligible.

We assume the photometric major axis is aligned with the axis of
maximum rotation, as required for an axisymmetric model. We set the
observed ellipticity of the Gaussian components to be $\epsilon=0.05$,
adjust the luminosity of the Gaussians accordingly to conserve flux,
and take the photometric major axis at PA=$198^\circ$ north through
east (Gebhardt et al.\ 1997). The chosen ellipticity sets the minimum
possible value of the inclination of M15 to be $25^\circ$, as
otherwise the Gaussians cannot be projected to have the observed
flattening. V05 showed that modest radial variations (or errors) in
the ellipticity are not critical for the resulting best-fit models.

\subsection{Aperture binning}
\label{sec:apbinning}

We fit the dynamical models to the observed kinematic data, binned in
apertures on the sky.  Since the models are axisymmetric, we first
reflect all the measurements to one quadrant, as described in V05, and
then construct a grid of polar apertures. The apertures
contain 50 stars per bin on average, except for the center where the bins contain only 10 stars. This allows an accurate
measurement of the mean velocity and velocity dispersion (V05).  We
use different sets of apertures for the proper motions and the
line-of-sight velocities. The 28 apertures and resulting $\langle v
\rangle$ and $\sigma$ for the line-of-sight data are shown in Figure
\ref{fig:kinbestfit}, and given in Table 3. The line-of-sight
apertures cover a large radial extent, from $7''$ to $10'$.  The
average velocity $\langle v \rangle$ and $\sigma$ measurement error is
1.4 \kms\ and 1.0 \kms, respectively. The proper motion data is
distributed over 13 apertures, also shown in
Figure~\ref{fig:kinbestfit}, with the corresponding values listed in
Table 4. The average velocity $\langle v \rangle$ and $\sigma$
measurement error is 0.04 mas yr$^{-1}$ and 0.03 mas yr$^{-1}$
respectively. The apertures extend to $18\farcs7$ from the center.\looseness=-2

\subsection{The parameters}
\label{sec:parameters}

M15 is a dense and old globular cluster in which substantial mass
segregation has taken place (e.g., Dull et al.\ 1997) so that $M/L$ is
expected to vary with radius. Our Schwarzschild models therefore have
not only the distance $D$ and inclination $i$ of M15 as free
parameters, but must allow for a radial $M/L$ variation.

In a constant $M/L$ model, the gravitational potential is obtained by
multiplying the luminosity of all the Gaussians in the MGE mass model
(Table 2) with the same $M/L$ value. To construct a mass model with a
smooth radial $M/L$ profile we varied the $M/L$ of the individual
Gaussians as this allows efficient calculation of the corresponding
gravitational potential. However, to reduce the number of free
parameters and to enforce a continuous profile we varied the first,
third, sixth and tenth Gaussian and interpolated the other Gaussians
logarithmically. The Gaussians ten through fourteen were given the
same $M/L$ value as Gaussian ten, because their individual $M/L$'s are
not constrained well because only line-of-sight velocities are
available at these radii. Finally, we include a central dark mass
$M_{\rm dark}$, represented by a point-mass potential.  As a result,
we have seven parameters: $D$, $i$, four for the $M/L$ profile, and
$M_{\rm dark}$, for which we construct models.

\subsection{Best-fit model}
\label{sec:bestfit}

Our dynamical models each have 2058 orbits covering a radial range of
$0\farcs16$ to $1923''$. Each model takes approximately 40 minutes on
an average 1.5Ghz desktop computer to complete. So a straightforward
search of the parameters is impractical as this would require a
minimum of $7^5=16807$ models.  Therefore we first searched the three
parameters $D$, $i$, $M_{dark}$ to find their best-fit values. After
that we searched $M_{dark}$ and the four $M/L$ parameters keeping $D$,
$i$, fixed at their best-fit values. Finally we checked that we found
the global minimum by doing a small search ($7^3$ models) through all
the parameters.

The model that best fits the photometric and kinematic observations of
M15 has a total of 275 constraints and a $\chi^2=88$. To determine the
error on the parameters we will use
$\Delta\chi^2=\chi^2-\chi^2_{min}$, where $\chi^2_{min}$ is the
$\chi^2$ of the best fitting model. We will use $\Delta\chi^2=3.53$
for the 68.3\% confidence for one free parameter for $D$ and $i$ and
$\Delta\chi^2=5.87$ for the 68.3\% confidence for 5 free parameters for
$M_{dark}$ and the $M/L$ profile. Since our reduced $\chi^2$ is much
smaller than one, our use of $\Delta\chi^2$ is conservative since we
likely over-estimate our uncertainties.

The best-fit model has the following parameters $D=10.3\pm0.4$ kpc ,
$i=60\pm15$, $M_{dark}=500^{+2500}_{-500} \Msun$ and a radially
varying $M/L$ profile. The best-fit $M/L$ values are tabulated in
Table 2 and shown in Figures~\ref{figmlprof} and
\ref{fig:chibhcnt} with their formal error bars. The best-fit
kinematics are shown in Figure~\ref{fig:kinbestfit}. Due to the small
number of stars in each aperture bin the scatter in the observed
kinematics is large. The proper motion mean velocities are dominated
by the errors. As a result, the mean proper motions of the best-fit
model are very small.  The overall rotation present in the
line-of-sight velocities is fitted with the model. Also the observed
dispersions are reproduced well.

\subsection{Inclination and Distance}
\label{sec:incdist}

\begin{figure}
\epsscale{1.0}
\plotone{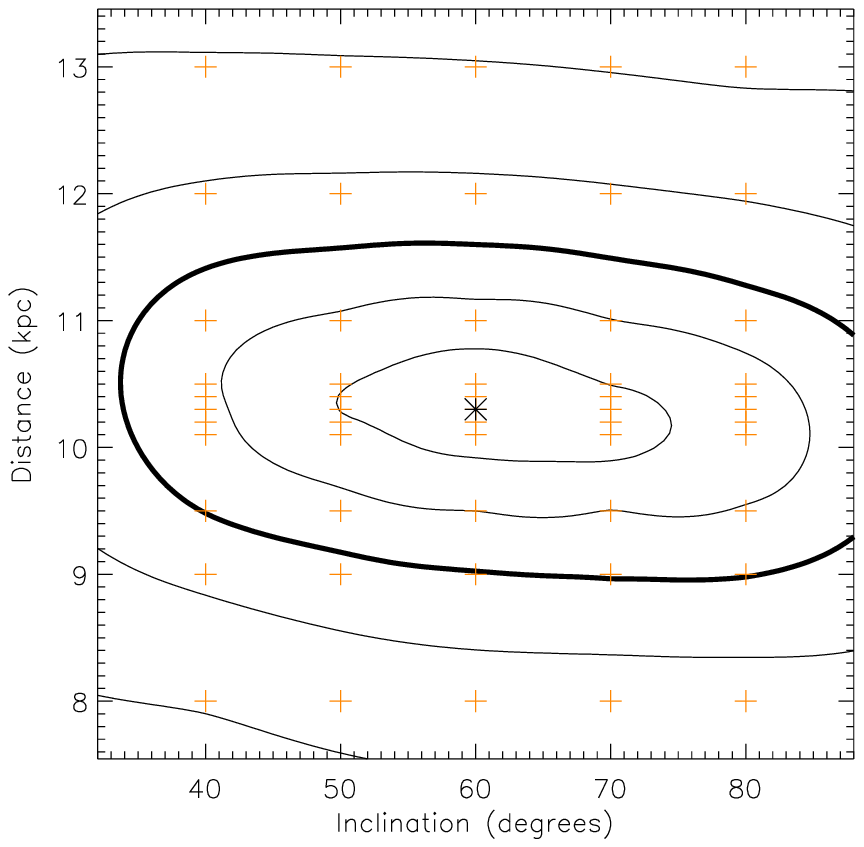}
\figcaption{\figchiincdist}
\end{figure}

The $D$ tan $i$ fit of Section~\ref{sec:glob-rot} gives an inclination
of $59^\circ\pm12^\circ$ at a distance of 10 kpc. The contours in
Figure~\ref{fig:chiincdist} show that the best-fitting inclination is
$60^\circ\pm15^\circ$ (68.3\% confidence, one parameters), and as such
the inclination is consistent with the $D$ tan $i$ prediction,
although it is not constrained very well.

The dynamical distance estimate of $10.0\pm0.5$ kpc from M04 which
assumes M15 is an isotropic sphere is very similar to our best-fit
distance. Our (consistent) distance is slightly larger, because we allow our models the freedom to be flattened, and have not restricted the
distribution function to be isotropic.

Our best-fit distance of $10.3\pm0.4$ kpc (68.3\% confidence, one
parameter) is in agreement with all other distance estimates:
$10.4\pm0.8$ kpc by Durrel \& Harris (1993), 10.3 kpc from the
globular cluster catalog of Harris (1996), $9.5\pm0.6$ kpc by
Silberman \& Smith (1995), and the Fe II metallicity-scale distance
11.2 kpc by Kraft \& Ivans (2003).

\begin{figure}
\epsscale{1.0}
\plotone{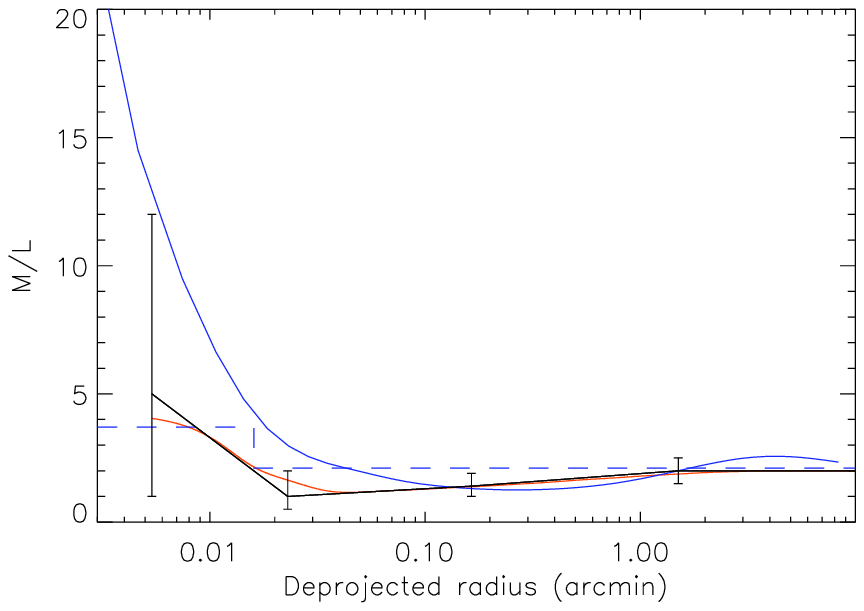}
\figcaption{\figmlprof}
\end{figure}

\subsection{The $M/L$ profile}
\label{sec:mlprofile}

Our best-fit $M/L$ model profile is shown in black in
Figure~\ref{figmlprof} with the formal error bars.  The error bars in
Figure~\ref{figmlprof} denote 68\% confidence errors of the five
parameter fit. Interpreting the error bars is difficult, since the
error bars are a one-dimensional view of the five-dimensional
parameter space. The $M/L$ values (and their errors) at the different
radii are strongly correlated, since they represent enclosed mass. The 
values are listed in Table 2, and can be used to convert the individual
Gaussian luminosity profiles into density profiles. Summing these 
provides the density profile for M15. Division by the luminosity profile
then gives a smooth $M/L$ profile. This is shown in red.

Pasquali et al.\ (2004) find an $M_\odot/L_\odot=2.1$ at $7'$ and an
$M_\odot/L_\odot=3.7$ in the center using a luminosity and mass
function derived from NICMOS data. Their results agree with our $M/L$
profile as shown in Figure~\ref{figmlprof} with a blue dashed line.
Another independent measurement of the central $M/L$ by Phinney (1993)
using the acceleration of pulsars in M15 yield a lower limit of the
$M_\odot/L_\odot$ of 2.1 inside 0\farcm1 and is consistent with our
profile.

The blue curve in Figure~\ref{figmlprof} is the radial $M/L$ profile
of M15 from an N-body model constructed by Baumgardt (priv.\ comm.),
which is rather similar to that of Dull et al.\ (2003). It has a
central peak, caused by compact remnants, mostly massive white
dwarfs. The total number of these remnants that survive the cluster
evolution is difficult to estimate, as it depends on the fraction of
neutron stars that is retained in the cluster potential after
supernova explosion, which is believed to endow them with substantial
kick velocities (Hansen \& Phinney 1997; Pfahl, Rappaport \&
Podsiadlowski 2002).

The profile from Baumgardt is significantly different from our
best-fit model inside 0\farcm6. To be able to accurately determine the
difference we made models with the Baumgardt $M/L$ profile instead of
our own. These models did not include a dark central mass. The
$\Delta\chi^2$ value we find for the best-fit model with Baumgardts
$M/L$ profile is outside our formal 99.9\% confidence level of our
best-fit model. The distance and inclination found by using the
Baumgardt profile do not differ significantly from our best-fit
values.

For completeness, we tried models with a constant $M/L$. The best-fit
distance, inclination and central dark mass do not change
significantly. The best-fit constant $M/L$ found is $1.6\pm0.2$ $
M_\odot/L_\odot$. This is the same as found by Gerssen et al.\ (2002) and consistent with 1.7 $M_\odot/L_\odot$ found by
Gebhardt (1997). The $\Delta\chi^2$ value found indicates that our
constant $M/L$ value is consistent with our $M/L$ profile within
95.4\% (two sigma) confidence levels.

Finally, the total mass of our best-fit model is
$4.4\times10^5\Msun$. This is in agreement with $4.9\times10^5\Msun$
(Dull et al.\ 1997), $4.4\times10^5\Msun$ (M04) and
$4.6\times10^5\Msun$ (G00). Our estimate of the total mass is
sensitive to the largely unconstrained value of $M/L$ outside of $2'$,
as this region contains $\sim40\%$ of the mass of the cluster. 

\subsection{Central dark mass}
\label{sec:mldcm}

Previous studies by Peterson et al.\ (1989), Gebhardt et al.\ (1997)
and Gerssen et al.\ (2002) have argued for an intermediate mass black
hole (IMBH) of up to a few thousand $M_\odot$ in M15. However M03 and
Baumgardt et al.\ (2003) reported that they can construct N-body
models for M15 which do not require an IMBH. For a detailed review of
the history of this controversial subject, see M03 and van der Marel
(2004).

Our formal best-fit value of the dark central mass is
$500^{+2500}_{-500} \Msun$ $(\Delta\chi^2=1$). This mass estimate agrees with all the
earlier estimates of the IMBH from Gebhardt (1997) to Gerssen et al.\
(2002). Figure~\ref{fig:chibhcnt} shows that there is a degeneracy between
the central $M/L$ and dark central mass. 

\begin{figure}
\epsscale{1.0}
\plotone{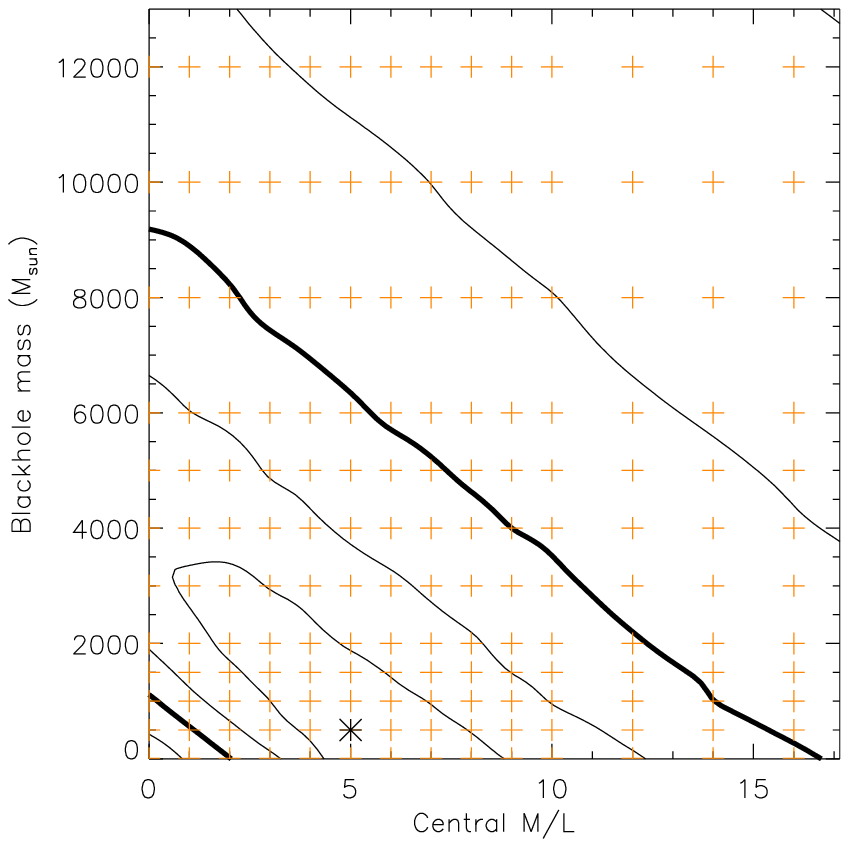}
\figcaption{\figchibhcnt}
\end{figure}

To be able to determine the inner structure, accurate information on
the central region is critical as it is difficult to determine what
happens in the central part that is not covered by luminosity and
kinematic data. The sphere of influence for a 1000 $M_\odot$ black
hole is $0\farcs5$. The inner luminosity data point that we used in
this model is at $0\farcs35$ from the center and the closest star is
$0\farcs4$ and $0\farcs25$ from the center for the line-of-sight and
proper motion data set, respectively. This results in a degeneracy
between the central $M/L$ and dark central mass as is shown in
Figure~\ref{fig:chibhcnt}. As a result we are not able to determine the
nature of the matter inside $\sim$1\farcs. By combining the mass of
our best-fit mass model and the dark central mass we estimate that the
total mass inside $1\farcs0$ arcsec (0.05 parsec, $10^4$ AU at 10.3
kpc) is 3400 $M_\odot$. This implies an extremely high central density
of 7.4 $\times 10^6$ M$_\odot
\pc^{-3}$.

To test the robustness of this density estimate we made models with a
constant $M/L$ (section \ref{sec:mlprofile}) and a dark central
mass. In this case we found the best-fit dark central mass to be
$1000^{+4000}_{-1000}$ $\Msun$. The mass contained inside $1\farcs0$ in
this model is comparable to that from our models with the best-fit
$M/L$ profile.

Guhathakurta et al.\ (1996) studied the cluster photometry of M15
using star counts. They detected 205 stars brighter than 20th
magnitude in V-band inside $1\farcs0$. The typical mass of these
post-main-sequence and turnoff stars is 0.75 $M_\odot$, thus these
stars account for 150 $M_\odot$. Their data is not able to constrain
the mass of fainter stars. But they give a rough (over) estimate and find
that the total stellar mass is 4000 $M_\odot$. When projected onto the
sky our mass model gives a total mass of 5000 $M_\odot$ in this
region. This would mean that there is room for dark mass in the form
of an IMBH with a mass in de order of 1000 $M_\odot$. If mass segregration has indeed taken place, the stellar
mass function used here might significantly overestimate the amount of low mass stars, and therefor lower the total stellar mass. This would allow for more dark matter inside $1\farcs0$.  

\begin{figure}
\plotone{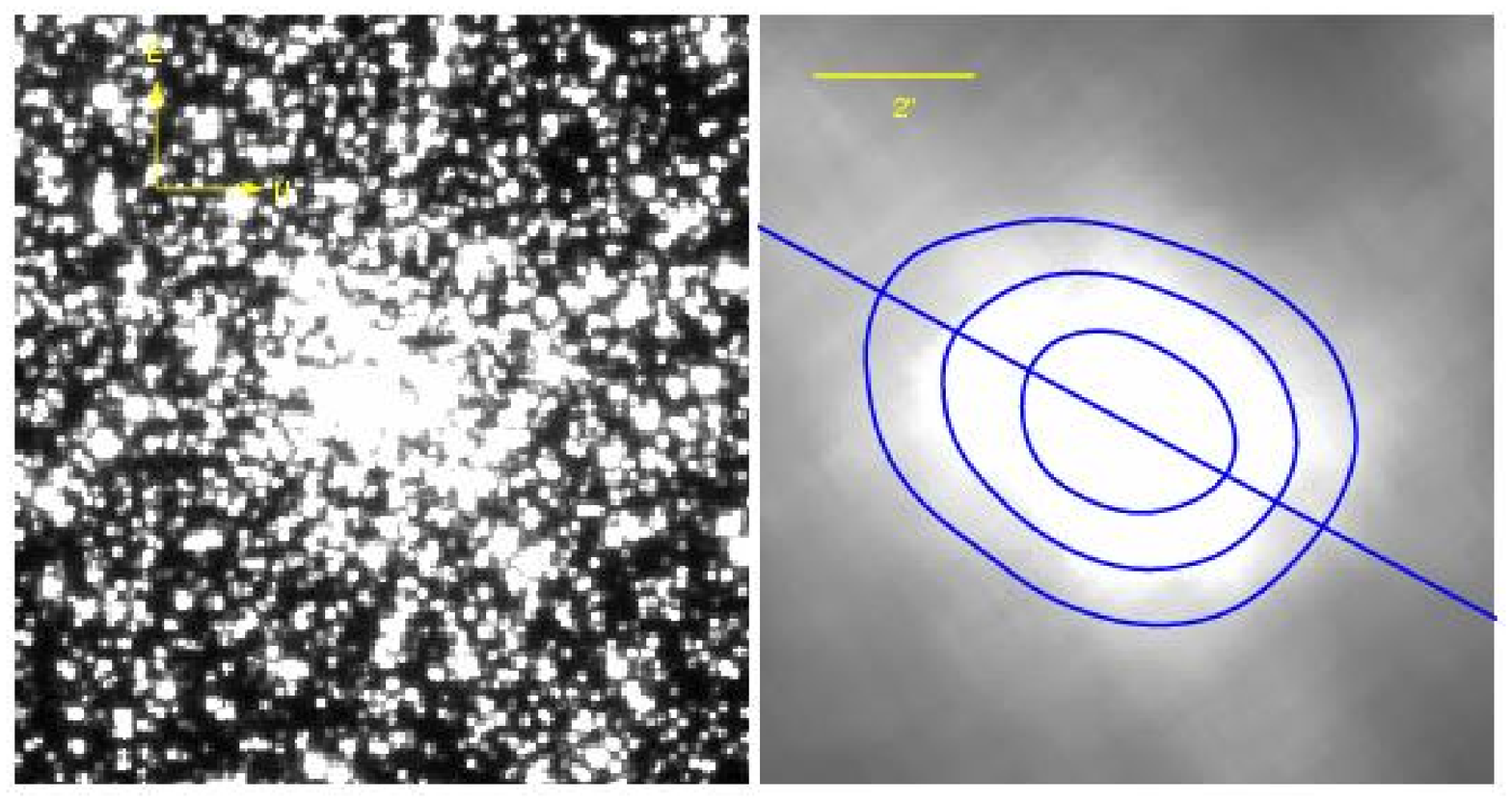}
\epsscale{1.0}
\figcaption{\figimage}
\end{figure}

\begin{figure}
\epsscale{1.0}
\plotone{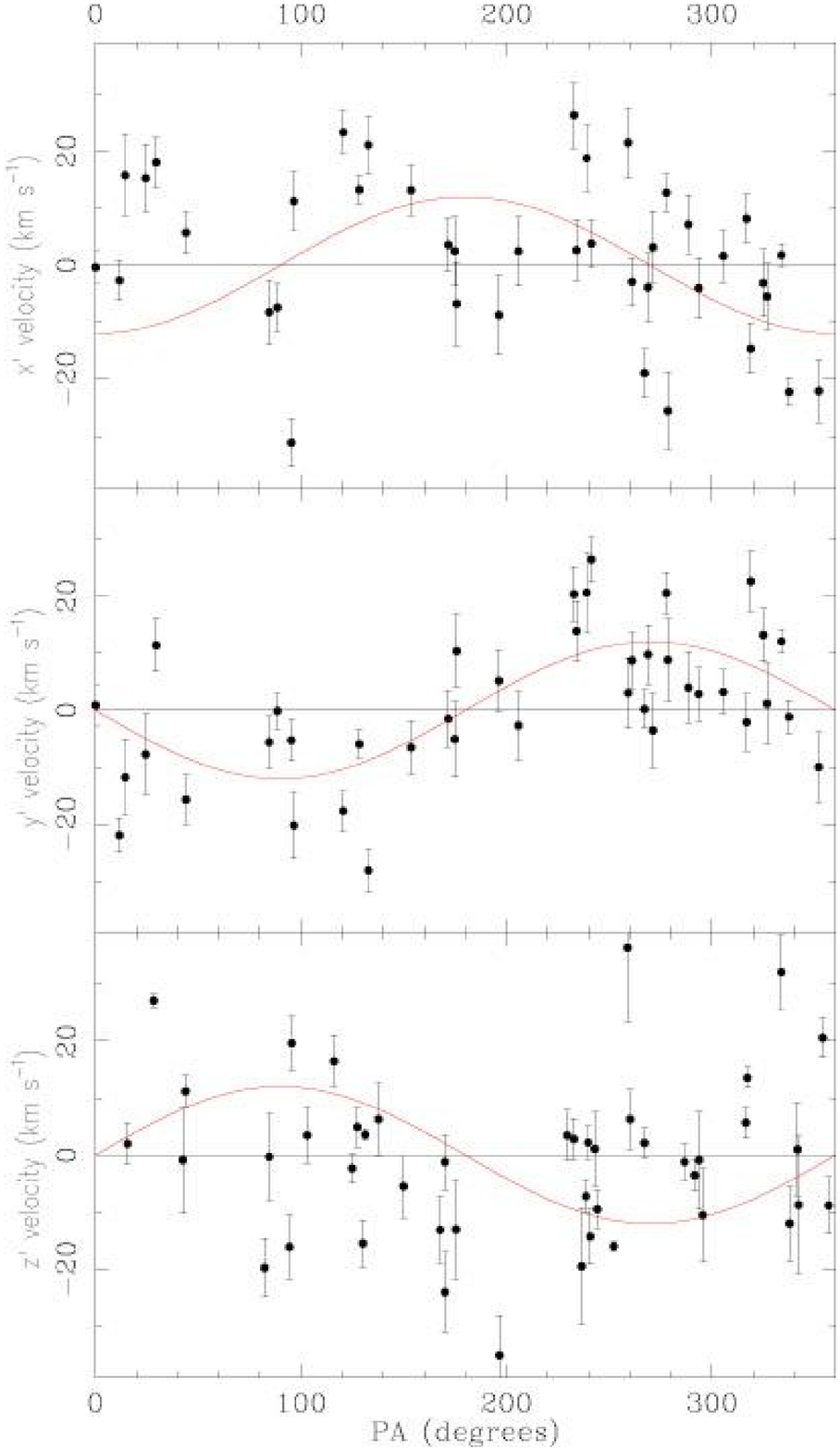}
\figcaption{\figrotpmrv}
\end{figure}

\section{A Decoupled Core?}
\label{sec:spinning-core}

The maps of the velocity fields shown in Figure~\ref{fig:mpmvel}
display significant structure in the inner few arcseconds. If M15 is
rotating, the plot of velocity against position angle for an annulus
will show a sinuisoidal variation (it is exactly sinusoidal for
rotation on cylinders). Figure~\ref{fig:rotpmrv} shows the individual
measurements in an annulus with inner and outer radii of $0\farcm03$ and
$0\farcm06$ as a function of position angle on the plane of the sky (cf.\
Figure 14 of G00). There is little azimuthal variation in $v_{x'}$,
but both $v_{y'}$ and $v_{z'}$ show a sinusoidal variation. Using
relation (\ref{e:dtani}), the line-of-sight velocity and $y'$ proper motion
suggest an inclination of $45\pm20^\circ$, consistent with our
dynamical modeling result. Since the proper motions and radial
velocites come from different dataset, the concordance of both results
and individual significance of the rotation in each component strongly
suggest that the central rotation in M15 is real. A simultaneous fit
to the two kinematic data sets, subject to the constraint
(\ref{e:dtani}), sets the $y'$-amplitude to $11\pm1.5$~\kms\ and
$x'$-amplitude to $7\pm1.5$~\kms, at a PA of $270\pm10^\circ$.

Figure~\ref{fig:image} shows the central region of M15 as observed
with WFPC2/F336W. We have heavily smoothed the image on the right-hand
side in order to measure any potential flattening. The image does show
substantial flattening, which appears in a F555W image as well. The
position angle of the major axis is not at either the kinematic PA nor
at the isophotal PA of the major axis at large radius.  We find a
position angle of $\sim 120^\circ$ (North to East) for the light
inside of $3''$. This PA is 30 degrees different from the kinematical
PA as defined by the rotation seen in the core. Although this
difference is significant given the uncertainty on the kinematically
defined PA, the photometrically determined PA is subject to
collections of bright stars and it is difficult to determine
uncertainties for it. Thus, we are unable to determine whether the
difference in the kinematic and photometric PA in the central region
is significant.

It is tempting to identify this misaligned structure with the
signature of an inspiraling binary IMBH, as described by Mapelli et
al.\ (2005). Their simulations of binary IMBH predict a small number of
stars with high rotational velocities close to the black hole, and
then a larger number of stars with aligned angular momentum (i.e.,
central rotation). However, we see no evidence for a small number of
associated high velocity stars, which would be a direct signature in
their scenario. We do though see ordered rotation. This rotation is
not direct evidence for a binary black hole, but it does suggest an
unexpected dynamical state for the central regions in M15. Since the
relaxation time is very short in the center of M15 (around $10^7$
years), rotation will be quickly removed by two-body interaction
(Akiyama \& Sugimoto 1989; Kim, Lee \& Spurzem 2004). As discussed in
Gebhardt et al. (2000), it is difficult to maintain such strong
rotation in the central parts of M15. A binary IMBH does increase the
central relaxation time (by lowering the stellar density) and offers a
possible formation mechanism---and even explains the misalignment between
the core and halo PA---but one would need stronger evidence in
order to invoke such a scenario.

The misaligned core is inconsistent with our assumption of an
axisymmetry dynamical model for M15, with major axis at PA of
$198^\circ$. When we ignore the kinematic measurements inside $4''$ in
the fitting procedure, we obtain the same distance and inclination.

\section{Discussion and Conclusions}
\label{sec:discussion}

We studied the globular cluster M15 by fitting line-of-sight
velocities, HST proper motions and surface brightness profiles with
orbit-based axisymmetric dynamical models.

The observations used for the modeling consisted of a luminosity
profile from Noyola \& Gebhardt (2005), 1264 line-of-sight velocity
measurements from G00 and a sample of 703 HST proper motions from
M03. The line-of-sight data extends out to $7'$, while the proper
motions cover the inner $0\farcm25$. The models provide a good fit to
the observations and allow us to measure the distance and inclination
of the cluster, the orbital structure, and the mass-to-light ratio
$M/L$ as a function of radius. We obtain a best-fit value for the
inclination of $i=60^\circ\pm15$ and a dynamical distance of
$D=10.3\pm0.4$ kpc, in good agreement with the canonical value. 

Our best-fit model has a $500^{+2500}_{-500}$ $\Msun$ dark central
mass and the $M/L$ profile shown in Figure~\ref{figmlprof}, which has
a central peak and a minimum at $0\farcm1$. The overall shape of the
profile resembles the shape expected for an expanded core globular
cluster (Dull et al.\ 2003). The central $M/L$ peak and the dark
central mass together represent a mass of 3400 $M_\odot$ inside the
inner $1\farcs0$ (0.05 parsec at a distance of 10.3 kpc).  This
suggests that the center harbors a large amount of dark mass. We
cannot distinguish the nature of the central mass concentration. It
could be an IMBH or it could be large number of compact objects, or it
could be a combination.

We found that a heavily smoothed image of M15 shows a flattened
structure inside $4''$, with a different PA from the outer part of the
cluster. The line-of-sight and proper motion data inside this radius
can be fitted using relation (1). The fit gives a PA similar to that
of the flattened structure and a rotational velocity of 10 \kms. This
suggests the structure is real, and constitutes a fast-spinning
decoupled core at the center of the cluster.

A significant improvement in the accuracy of the dynamical models for
M15 is possible by increasing the accuracy of the proper motions and
radial velocities.  This appears possible with the ACS onboard HST,
and with high-resolution spectrographs on 8m class telescopes.\looseness=1

\acknowledgments


KG acknowledges upport from NASA through grant number HST-AR-09542
from the Space Telescope Science Institute, which is operated by AURA,
Inc., under NASA contract NAS5--26555 and NSF CAREER grant
AST-0349095. TdZ gratefully records the warm hospitality of Neal Evans
and Leslie Geballe during part of this work.\looseness=-2



\begin{deluxetable}{rrrrrrrrrrrrrr}
\tablewidth{16truecm}
\tablecaption{Kinematics of the proper motions in polar 
               apertures\label{tab:propdata}}
\tablehead{
  \colhead{\#} & \colhead{$n_\star$} & \colhead{$r_0$}
& \colhead{$\theta_0$} & \colhead{$\Delta r$} 
& \colhead{$\Delta\theta$} & \colhead{$V_{x'}$} 
& \colhead{$\Delta V_{x'}$} & \colhead{$\sigma_{x'}$} 
& \colhead{$\Delta\sigma_{x'}$} & \colhead{$V_{y'}$}
& \colhead{$\Delta V_{y'}$} & \colhead{$\sigma_{y'}$} 
& \colhead{$\Delta\sigma_{y'}$} \\
\colhead{(1)} & \colhead{(2)} & \colhead{(3)} & \colhead{(4)} &
\colhead{(5)} & \colhead{(6)} & \colhead{(7)} & \colhead{(8)} &
\colhead{(9)} & \colhead{(10)} & \colhead{(11)} & \colhead{(12)} &
\colhead{(13)} & \colhead{(14)} \\[-10pt]
}
\startdata
 1&50&0.021&0.785&0.033&1.571&-0.04& 0.04& 0.25& 0.03&-0.02& 0.04& 0.24&0.04\\
 2&51&0.060&0.393&0.046&0.785&-0.02& 0.04& 0.21& 0.02&-0.02& 0.04& 0.22&0.03\\
 3&54&0.060&1.178&0.046&0.785&-0.04& 0.03& 0.23& 0.02&-0.01& 0.04& 0.25&0.03\\
 4&54&0.108&0.262&0.050&0.524& 0.01& 0.04& 0.28& 0.02&-0.00& 0.03& 0.19&0.03\\
 5&54&0.108&0.785&0.050&0.524& 0.06& 0.04& 0.22& 0.03& 0.07& 0.03& 0.19&0.03\\
 6&63&0.108&1.309&0.050&0.524& 0.02& 0.03& 0.23& 0.03& 0.04& 0.03& 0.17&0.03\\
 7&45&0.154&0.262&0.043&0.524& 0.02& 0.04& 0.22& 0.02& 0.04& 0.04& 0.26&0.03\\
 8&48&0.154&0.785&0.043&0.524&-0.06& 0.03& 0.21& 0.03&-0.01& 0.03& 0.15&0.03\\
 9&61&0.154&1.309&0.043&0.524& 0.03& 0.03& 0.19& 0.02&-0.02& 0.02& 0.19&0.02\\
10&56&0.214&0.262&0.077&0.524&-0.02& 0.02& 0.15& 0.03&-0.02& 0.03& 0.22&0.02\\
11&58&0.214&0.785&0.077&0.524& 0.02& 0.03& 0.21& 0.03& 0.00& 0.03& 0.25&0.03\\
12&62&0.214&1.309&0.077&0.524&-0.05& 0.03& 0.18& 0.02&-0.01& 0.03& 0.25&0.03\\
13&47&0.279&0.785&0.053&1.571& 0.01& 0.04& 0.21& 0.02&-0.03& 0.04& 0.24&0.02\\

\enddata 
\tablecomments{
The mean velocity and velocity dispersion of the proper motion
observations calculated in polar apertures on the plane of the sky.
Per row the information per aperture is given. The first column labels
the aperture and the second column gives the number of stars $n_\star$
that fall in the aperture. Columns 3--6 list the polar coordinates $r$
(in arcmin) and the angle $\theta$ (in degrees) of the centroid of the
aperture and the corresponding widths $\Delta r$ (in arcmin) and
$\Delta\theta$ (in degrees). The remaining columns present the average
proper motion kinematics in units of \masyr. The mean velocity $V$
with error $\Delta V$ and velocity dispersion $\sigma$ with error
$\Delta\sigma$ are given in columns 7--10 for the proper motion
component in the $x'$-direction and in columns 11--14 for the proper
motion component in the $y'$-direction. }
\end{deluxetable}

\begin{deluxetable}{rrrrrrrrrr}
\tablewidth{13.0truecm}
\tablecaption{Kinematics of the line-of-sight velocities in polar 
               apertures\label{tab:vlosdata}}
\tablehead{
  \colhead{\#} & \colhead{$n_\star$} & \colhead{$r_0$}
& \colhead{$\theta_0$} & \colhead{$\Delta r$} 
& \colhead{$\Delta\theta$} & \colhead{$V_{z'}$} 
& \colhead{$\Delta V_{z'}$} & \colhead{$\sigma_{z'}$} 
& \colhead{$\Delta\sigma_{z'}$} \\
\colhead{(1)} & \colhead{(2)} & \colhead{(3)} & \colhead{(4)} &
\colhead{(5)} & \colhead{(6)} & \colhead{(7)} & \colhead{(8)} &
\colhead{(9)} & \colhead{(10)}  \\[-10pt]
}
\startdata

 1 & 10 & 0.015 & 0.785 & 0.022 & 1.571 &  0.6 &  3.8 & 10.8 &  2.6 \\
 2 & 10 & 0.030 & 0.785 & 0.008 & 1.571 &  1.0 &  4.1 & 10.4 &  2.5 \\
 3 & 10 & 0.040 & 0.785 & 0.011 & 1.571 & -0.7 &  4.8 & 13.0 &  3.8 \\
 4 & 10 & 0.049 & 0.785 & 0.007 & 1.571 & -8.9 &  4.2 & 11.2 &  2.2 \\
 5 & 62 & 0.103 & 0.393 & 0.102 & 0.785 &  0.2 &  1.6 & 11.3 &  1.1 \\
 6 & 56 & 0.103 & 1.178 & 0.102 & 0.785 &  1.2 &  1.8 & 12.8 &  1.4 \\
 7 & 55 & 0.214 & 0.262 & 0.120 & 0.524 &  1.4 &  1.4 &  9.4 &  0.9 \\
 8 & 63 & 0.214 & 0.785 & 0.120 & 0.524 &  2.3 &  1.3 &  9.3 &  1.0 \\
 9 & 52 & 0.214 & 1.309 & 0.120 & 0.524 &  3.2 &  1.4 &  9.4 &  0.9 \\
10 & 63 & 0.366 & 0.196 & 0.184 & 0.393 &  0.6 &  1.4 & 10.3 &  1.0 \\
11 & 63 & 0.366 & 0.589 & 0.184 & 0.393 &  0.5 &  1.4 & 10.3 &  1.0 \\
12 & 54 & 0.366 & 0.982 & 0.184 & 0.393 &  1.8 &  1.9 & 12.3 &  1.1 \\
13 & 70 & 0.366 & 1.374 & 0.184 & 0.393 &  1.7 &  1.2 & 10.2 &  0.9 \\
14 & 53 & 0.553 & 0.196 & 0.190 & 0.393 &  0.7 &  1.5 & 10.2 &  1.2 \\
15 & 57 & 0.553 & 0.589 & 0.190 & 0.393 &  3.6 &  1.5 &  9.8 &  0.8 \\
16 & 63 & 0.553 & 0.982 & 0.190 & 0.393 &  2.5 &  1.2 &  9.6 &  0.8 \\
17 & 60 & 0.553 & 1.374 & 0.190 & 0.393 &  0.7 &  1.3 &  9.5 &  0.8 \\
18 & 61 & 0.791 & 0.196 & 0.286 & 0.393 &  2.7 &  1.3 &  9.0 &  0.8 \\
19 & 69 & 0.791 & 0.589 & 0.286 & 0.393 &  2.5 &  1.2 &  9.3 &  0.8 \\
20 & 78 & 0.791 & 0.982 & 0.286 & 0.393 &  2.9 &  1.1 &  8.5 &  1.0 \\
21 & 48 & 0.791 & 1.374 & 0.286 & 0.393 &  3.4 &  1.5 &  9.9 &  2.1 \\
22 & 68 & 1.165 & 0.196 & 0.462 & 0.393 &  3.4 &  1.0 &  7.8 &  0.8 \\
23 & 49 & 1.165 & 0.589 & 0.462 & 0.393 &  1.0 &  1.3 &  8.0 &  1.2 \\
24 & 66 & 1.165 & 0.982 & 0.462 & 0.393 &  0.6 &  1.1 &  8.5 &  1.4 \\
25 & 62 & 1.165 & 1.374 & 0.462 & 0.393 &  1.2 &  1.2 &  9.1 &  2.0 \\
26 & 60 & 2.560 & 0.393 & 2.329 & 0.785 &  1.8 &  0.8 &  6.5 &  0.7 \\
27 & 66 & 2.560 & 1.178 & 2.329 & 0.785 &  0.4 &  0.9 &  6.7 &  1.0 \\
28 & 60 & 5.306 & 0.785 & 3.163 & 1.571 &  1.5 &  0.6 &  3.9 &  0.4 \\
\enddata
\tablecomments{
The mean velocity and velocity dispersion calculated in polar
apertures on the plane of sky from the line-of-sight velocity
observations. Columns 1--6 are as in Table~\ref{tab:propdata} and the
remaining columns present the average line-of-sight kinematics in
\kms.  }
\end{deluxetable}

\clearpage



\end{document}